\documentstyle[12pt]{article}
\begin{document}
\begin{center}
\section*{\bf Stringent limits on the existence of Planck time from
stellar interferometry}
\vspace{1.5mm}

Richard Lieu and Lloyd W. Hillman
 
Department of Physics, University of Alabama, Huntsville,
AL 35899, U.S.A.
\end{center}
 
\vspace{1.5mm}
 
\noindent
{\bf Abstract}

We present a method of directly testing whether time is `grainy'
on scales of $\leq t_P = (\hbar G/c^5)^{\frac{1}{2}} \approx
5.4 \times 10^{-44}$ s, the Planck time.  If the phenomenon exists, the
energy and momentum of a photon (quantities which are independently 
measurable) will be subject to ultimate uncertainties of the form 
$\delta E/E \approx \delta p/p \approx (E/E_P)^\alpha$, where $E_P = h/t_P$ 
and $\alpha \sim 1$, because surpassing these will lead to `super-clocks' and
`super-rulers'.  A well-known consequence is random perturbation of the photon
dispersion equation by a correction term which to lowest order modifies the
equation to the form $p^2 = E^2 [1 \pm a_o (E/E_P)^\alpha]$, where $a_o \sim 1$
and $c=1$.  As a result, the two wave 
velocities will fluctuate by different amounts, so that after propagating a 
sufficiently large distance the phase of the radiation will no longer be 
well-defined, however sharp it might have been when emitted by the source.  
Since, at optical frequencies, the technique of stellar interferometry readily 
ascertains whether light from an astronomical object at various distances 
retains its phase information upon arrival, it is shown here that such 
observations place stringent limits on the graininess of time.  Currently the 
furthest star from which interference fringes were seen, S Ser, is at $\sim$
1 kpc, implying that all models with $\alpha \leq \frac{13}{15}$ are to be 
rejected, including scenarios based upon random-walk ($\alpha = 1/2$) and the 
holographic principle of Wheeler and Hawking ($\alpha = 2/3$),  The decisive 
step, of course, is to test the presence or not of the first order term 
$E/E_P$ itself (i.e. $\alpha =1$); for this one must await the imminent 
observations of extra-galactic sources at $> 1$ Mpc by the Very Large 
Telescope Interferometer (here-and-after the VLTI).  Based upon the S Ser 
result, nonetheless, it is already possible to conclude that there is no first
order departure from exactness in time over intervals $\approx 4.2 \times 
10^{-40}$ s ($\approx 10^4$ times longer than the Planck time).

\newpage

It is widely believed that time ceases to be exact
at intervals $\leq t_P$,
where quantum fluctuations
in the vacuum metric tensor renders General Relativity an
inadequate theory.  Both $t_P$ and its corresponding distance
scale $l_P = c t_P$, the Planck length, play a vital role in almost
all of the many theoretical models (including superstrings) that
attempt to explain how the universe was born, and how it evolved during
infancy (see e.g, Silk 2000 and references therein).  Unfortunately we lack
experimental evidence for the existence or not of
$t_P$.  Although the recent efforts in using data from
gravitational wave interferometry and the observation of
ultra-high energy (UHE) quanta carry potential
(Amelino-Camelia 2001, Ng et al 2001,
Lieu 2002),
they are still some way from delivering a verdict.
Here we wish to describe how an entirely different
yet well established technique has hitherto been overlooked: not only would it
enable direct tests for Planck scale fluctuations
(and revealing the detailed properties of any such effects),
but also the measurements performed to date could already be used to
eliminate prominent theories.

Owing to the variety of proposed models we begin by describing the
common feature
that define the phenomenon being searched:
if  a time $t$ is so small that $t \rightarrow t_P$
even the best clock ever made will only be able to determine it
with an uncertainty $\delta t \geq t$.  To express this
mathematically we may write the intrinsic
standard deviation of time as $\sigma_t/t = f(t_P/t)$, where $f \ll 1$ for
$t \gg t_P$ and $f \geq 1$ for $t \leq t_P$.  Over the range $t \gg t_P$
the (hitherto unknown) function $f$ can be expanded as follows:
\begin{equation}
f(x) = x^\alpha (a_o + a_1 x + a_2 x^2 + ...) \approx a_o x^\alpha~
{\rm for}~x \ll 1,~
{\rm where}~x = \frac{t_P}{t}~{\rm and}~a_o \sim 1,
\end{equation}
and $\alpha$ is an index which in principle can assume any positive value.
Since for the rest of this paper we shall be concerned only with
times $t \gg t_P$  we may take an approximate form of Equ. (1) as:
\begin{equation}
\frac{\sigma_t}{t} \approx \left(\frac{t_P}{t}\right)^\alpha
\end{equation}

Our appreciation of how
Equ. (2) may affect measurements
of $(E, {\bf p})$  arises
from the realization that if frequencies $\nu > \nu_P = 1/t_P$
can be determined accurately such a calibration will lead to a
`superclock' that keeps time to within $\delta t < t_P$.  Thus it
should be possible to demonstrate that as $\nu \rightarrow \nu_P$,
$\delta \nu/\nu \rightarrow 1$.  In fact,
for the case of $\sigma_t \approx t_P$ (i.e. Equ. (2) 
with $\alpha =1$) the following Equ. was shown by Lieu (2002)
to be an immediate consequence:
\begin{equation}
\frac{\sigma_{\nu}}{\nu}
\approx \frac{\nu}{\nu_P},~{\rm or}
\frac{\sigma_{E}}{E} \approx \frac{E}{E_P},
\end{equation}
where $E = h \nu$ and $E_P = h \nu_P =
h/t_P \approx 8.1 \times 10^{28}$ eV.
Further, for any value of $\alpha$ it can be proved 
(see Ng \& van Dam 2000) that Eq. (2)
leads to:
\begin{equation}
\frac{\sigma_{\nu}}{\nu}
\approx \left(\frac{\nu}{\nu_P}\right)^\alpha,~{\rm or}~
\frac{\sigma_{E}}{E} \approx \left(\frac{E}{E_P}\right)^\alpha.
\end{equation}
The same reasoning also applies to the
intrinsic uncertainty in data on the momentum
${\bf p}$ (note however for measurements directly taken by an observer
$\delta E$ and $\delta {\bf p}$,
like $\delta t$ and $\delta {\bf r}$,
are uncorrelated errors), for
if any component of ${\bf p}$ could be known to high
accuracy even in the limit of large $p$ we would be able to surpass
the Planck length limitation in distance determination for that direction.
Thus a similar equation may then be formulated as:
\begin{equation}
\frac{\sigma_p}{p} \approx \left(\frac{E}{E_P}\right)^\alpha
\end{equation}
where $p$ is the magnitude of ${\bf p}$ and the right side is identical to the
previous equation because $E \approx p$ for photons.
Note indeed that
Equs. (4) and (5) hold good for ultra-relativistic particles as well.

About the value of $\alpha$, the straightforward choice
is $\alpha = 1$, which by Equ. (2) implies $\sigma_t \approx t_P$,
i.e. the most precise clock has uncertainty $\sim t_P$.  Indeed,
$\alpha = 1$ is just the first order term in a power series expansion of
quantum loop gravity.  However, the quantum nature of time at scales
$\leq t_P$ may be
manifested in  other (more contrived) ways.
In particular,  for random walk models of space-time, where
each step has size $t_P$, $\alpha = 1/2$ (Amelino-Camelia 2000).
On the other hand, it was shown
(Ng 2002)
that as a consequence of
the holographic principle
(which states that the maximum degrees of freedom allowed within a
region of space is given by the volume of the region in Planck units, see
Wheeler 1982, Bekenstein 1973, Hawking 1975, 't Hooft 1993, Susskind 1995)
$\sigma_t$ takes the form $\sigma_t/t = (t_P/t)^{\frac{2}{3}}$, leading
to $\alpha = 2/3$ in Eqs. (3) and (4).  Such an undertaking also
has the desirable property  (Ng 2002)
that it readily implies a finite lifetime
$\tau$ for
black holes, viz.,  $\tau \sim G^2 m^3/\hbar c^4$, in agreement with
the earlier calculations of Hawking.  Although the choice of $\alpha$
is not unique, the fact that it appears as an exponent means
different values can lead to wildly varying predictions.  Specifically,
even by taking $E = 10^{20}$ eV (i.e. the highest energy particles known,
where Planck scale effects are
still only $\sim (E/E_P)^\alpha \approx 10^{-9 \alpha}$ in
significance),
an increment of $\alpha$ by 0.5  would demand a
detection sensitivity 4.5 orders of
magnitude higher.  The situation gets much worse as $E$ becomes lower.
Thus if an experiment fails to offer confirmation at a
given $\alpha$, one can always raise the value of $\alpha$, and
the search may never end.  Fortunately, however, it turns out that
all of the three scenarios $\alpha = 1/2, 2/3$, and $1$ may now be
observationally clinched.

How do Equs. (4) and (5) modify our perception of the photon dispersion
relation?  By writing the relation as: 
\begin{equation}
E^2 - p^2 = 0
\end{equation}
the answer becomes clear - one simply needs to calculate the uncertainty
in $E^2 - p^2$ due to the 
intrinsic fluctuations in the measurements of $E$ and ${\bf p}$, viz.
$\delta (E^2 - p^2) =
2E \delta E - 2p \delta p$, bearing in mind that $\delta E$ and $\delta p$
are independent variations, as already discussed.  This allows us
to obtain the standard deviation
\begin{equation}
\sigma_{E^2-p^2} =
2 \sqrt{2} E^2 \left(\frac{E}{E_P}\right)^\alpha
\end{equation}
Thus, typically a photon will go off-shell, with Eq. (6) replaced by:
\begin{eqnarray}
E^2 - p^2 \approx \pm 2 \sqrt{2} E^2 \left(\frac{E}{E_P}\right)^\alpha
\end{eqnarray}
In the case of a positive fluctuation on the right hand term of Eq. (6)
by unit $\sigma$,
the phase and group velocities of propagation will read, for
$E/E_P \ll 1$, as:
\begin{eqnarray}
v_p = \frac{E}{p} \approx 1 + \sqrt{2} \left(\frac{E}{E_P} \right)^\alpha;~~
v_g = \frac{dE}{dp} \approx 1 + \sqrt{2} (1 + \alpha)
\left(\frac{E}{E_P}\right)^\alpha
\end{eqnarray}
The results differ from that of a particle - here $E^2 - p^2$
is a function of $E$ and not a constant, so that both
$v_p$ and $v_g$ are $> 1$, i.e. greater than the speed of light $v=1$.
On the other hand, if the right side of Eq. (6) fluctuates negatively
the two wave velocities will read like:
\begin{eqnarray}
v_p = \frac{E}{p} \approx 1 - \sqrt{2} \left(\frac{E}{E_P} \right)^\alpha;~~
v_g = \frac{dE}{dp} \approx 1 - \sqrt{2} (1 + \alpha)
\left(\frac{E}{E_P} \right)^\alpha
\end{eqnarray}
and will both be $< 1$.

Is it possible to force a re-interpretation of Equ. (8) in another (more
conventional) way, viz., for
a particular off-shell mode $E^2 - p^2$ typically assumes a 
{\it constant} value
different from zero by about the unit $\sigma$
of Equ. (7)?  The point, however, is that
even in this (very artificial)
approach, whereby the photon is regarded as a particle of
non-vanishing but fixed $m^2$, the quantities $v_p = E/p$ and $v_g = dE/dp =
p/E$ will still disagree with each other randomly by an amount 
$\sim (E/E_P)^\alpha$, so that the chief outcome of
Equs. (9) and (10) is robust.

But is such an effect observable?  Although an obvious approach is
to employ the highest energy radiation, so as to maximize $E/E_P$,
such quanta are difficult to detect.  More familiar types of
radiation, e.g. optical light
at $E \approx 1$ eV ($\lambda \approx
1.24 \mu m$), have much smaller values of $E/E_P$,
yet the advantage is that we can measure their properties with great accuracy.
Specifically we consider the phase behavior of 1 eV light
received from a
celestial optical source
located at a distance $L$ away.  During the 
propagation time $\Delta t = L/v_g$, the phase has advanced from its
initial value $\phi$ (which we assume to be well-defined)
by an amount:
\begin{eqnarray}
\Delta \phi = 2 \pi \frac{v_p \Delta t}{\lambda} = 2 \pi \frac{v_p}{v_g} 
\frac{L}{\lambda} \nonumber
\end{eqnarray}
According to Equs. (9) and (10), $\Delta \phi$ should then randomly fluctuate
in the following manner:
\begin{equation}
\Delta \phi =  2 \pi \frac{L}{\lambda} \left[1 \pm \sqrt{2} \alpha
\left(\frac{E}{E_P}\right)^\alpha \right]
\end{equation}
In the limit when
\begin{equation}
\sqrt{2} \alpha \left(\frac{E}{E_P}\right)^\alpha \frac{L}{\lambda} \geq 1;~~
{\rm or}~~ \frac{\sqrt{2} \alpha}{h} E^{1+\alpha} E_P^{-\alpha} L \geq 1
\end{equation}
the phase  of the wave will have appreciable
probability of assuming any value between $0$ and $2 \pi$ upon arrival, 
irrespective of how sharp the initial phase at the source may be.

From the preceding paragraph, a way towards directly
testing whether time remains exact at the Planck scale
has become apparent.  In stellar interferometry
(see e.g. Baldwin \& Haniff 2002 for a review)
two light rays from an astronomical source
travel along different paths to reach two reflectors (within
a terrestrial telescope) which subsequently coverge them to
form  interference
fringes.  By Equ. (11), however,  we see that if the time quantum exists
the phase of light from a 
sufficiently distant
source will appear random - when $L$ is large enough to satisfy Equ. (12)
the fringes will disappear.  In fact, 
the value of $L$ at which Equ. (12) holds may readily be calculated
for the case of $\alpha = 2/3$ and $\alpha = 1$,
with the results:
\begin{equation}
L \geq 2.47 \times 10^{15} (E/1~{\rm eV})^{-\frac{5}{3}} 
~~{\rm cm}~~(\alpha = \frac{2}{3});~~
L \geq 7.07 \times 10^{24} (E/1~{\rm eV})^{-2}
~~{\rm cm}~~(~~\alpha = 1).    
\end{equation}
These distances correspond respectively to 165 AU (or $8 \times 10^{-4}$ pc)
and 2.3 Mpc.   

It is interesting to note that interference effects were clearly seen
at $\lambda =$ 2.2 $\mu m$ ($E \approx$ 0.56 eV) light from the star S Ser
at 1.012 kpc, using the
Infra-red Optical Telescope Array, which enabled a radius determination of
the star (van Belle, Thompson, \& Creech-Eakman 2002).  When comparing
with Equ. (13) one realizes that this finding has already
excluded the possibility of  $\alpha = 2/3$, because for such a case
$\Delta \phi$ should carry uncertainties $\gg 2 \pi$, and the two
light rays would not have interfered.
Thus the presence of fringe patterns  from S Ser
implies an absence of 
the second (correction) term on the right side of
Equ. (11) for all $\alpha \leq \frac{13}{15}$.  
In another manner of expression,
the S Ser observation has set an upper limit on the effects of
Planck scale fluctuations at the level of $(E/E_P)^\frac{13}{15} =
5.42 \times 10^{-26}$ (or $\approx 2$ parts in $10^{25}$), while the
genuine first order correction term is at the level $\sim E/E_P
= 6.96 \times 10^{-30}$; the ratio of the two numbers gives, in
units of $t_P$, the smallest interval over which we currently
know that time must remain to first order a precise quantity.
This interval is $\approx 7.8
\times 10^3 t_P$, or $4.2 \times 10^{-40}$ s.

Obviously, the milestone point is not yet
reached until we can clinch the $E/E_P$ ($\alpha =1$) term
itself.  From Equ. (13)
we see that this crucially
await the observation of extragalactic objects, which can be done
in the forseeable future by the
VLTI of the European Southern
Observatory, as it carries an assembly of four 8m mirrors to provide
sensitivity for such measurements on sources like M87 and 3C273 
at 1-2.5  $\mu m$ wavelengths
(Richichi et al 2002).

RL is grateful to Jonathan Mittaz, Sir Ian Axford, and Lord James
McKenzie of the Hebrides and Outer Isles for discussions.

\noindent
{\bf References}

\noindent
Amelino-Camelia, G., 2000, Towards quantum gravity, Proc. of the XXXV
International\\
\indent Winter School on Theor.  Phys., Polanica, Poland, Ed. Jerzy Kowalski-
Glikman.\\
\indent Lecture Notes in Phys., 541, 1, Berlin: Springer-Verlag.\\
\noindent
Amelino-Camelia, G., 2001, Nature, 410, 1065.\\
\noindent
Baldwin, J.E., \& Haniff, C.A., 2002, Phil. Trans. A360, 969.\\
\noindent
Bekenstein, J.D., 1973, Phys. Rev. D., 7, 2333.\\
\noindent
Hawking, S., 1975, Comm. Math. Phys., 43, 199.\\
\noindent
t'Hooft, G., 1993, in {\it Salamfestschrift}, p. 284, Ed. A Ali et al
(World Scientific,\\ 
\indent Singapore).\\
\noindent
Lieu, R., 2002, ApJ, 568, L67.\\
\noindent
Ng, Y. -J., \& van Dam, H., 2000, Found. Phys., 30, 795.\\
\noindent
Ng, Y. -J., Lee, D. -S., Oh, M.C., \& van Dam, H., 2001, Phys. Lett, B507,
236. \\
\noindent
Ng, Y. -J., 2002, Int. J. Mod. Phys., D., in press (to appear in Dec.
special issue), gr-qc/0201022. \\
\noindent
Richichi, A., Bloecker, T., Foy, R., Fraix-Burnet, D., Lopez, B., Malbet, F.,
Stee, P.,\\
\indent von der Luehe, O., \\
Weigelt, G., 2000, Proc. SPIE, 4006, 80.\\
\noindent
Silk, J., 2001, The Big Bang, 3rd ed., W.H. Freeman \& Co.\\
\noindent
Susskind, L., 1995, J. Math. Phys (N.Y.), 36, 6377.\\
\noindent
van Belle, G.T., Thompson, R.R., Creech-Eakman, M.J., 2002, AJ, 124, 1706.\\
\noindent
Wheeler, J., 1982, Int. J. Theor. Phys., 21, 557. \\

\end{document}